\documentclass[aps,prl,twocolumn,superscriptaddress,showpacs]{revtex4}

\usepackage{epsfig}
\usepackage{graphics}
\usepackage{amsmath}
\usepackage{color}

\newcommand{\ket}[1]{\left\vert#1\right\rangle}

\newcommand{\Rb}{$^{87}Rb$ }
\newcommand{\fig}[1]{\mbox{Fig. #1}}
\newcommand{\ppsi}{\left\vert \psi \right\rangle}

\begin{document}
\author{Yoav Sagi}
\affiliation{Department of Physics of Complex Systems, Weizmann Institute of Science, Rehovot 76100, Israel }
\author{Ido Almog}
\affiliation{Department of Physics of Complex Systems, Weizmann Institute of Science, Rehovot 76100, Israel }
\author{Nir Davidson}
\affiliation{Department of Physics of Complex Systems, Weizmann Institute of Science, Rehovot 76100, Israel }
\title{Suppression of collisional decoherence}
\begin{abstract}
We employ a continuous dynamical decoupling scheme to suppress the decoherence induced by elastic collisions of cold atoms. Using a continuous echo pulse we achieve a thirty-fold increase in the coherence time of $^{87}Rb$ atoms trapped in a dipole trap. Coherence times of more than 100ms are demonstrated for an ensemble with an optical depth of $120$.
\end{abstract}
\maketitle

An ensemble of two level quantum systems coupled to a fluctuating external environment is a common paradigm in many fields of study. This coupling leads to decoherence that limits the usefulness of these systems. Application of external pulses can reduce the decoherence by utilizing symmetry properties of the coupling Hamiltonian to average out its effect, a method commonly referred to as Dynamical Decoupling (DD) \cite{Viola1998,Kofman2001}. Experimentally, DD is used in solid NMR \cite{Haeberlen_NMR_book_1976} and it was recently demonstrated to increase the coherence time of trapped ions \cite{Biercuk2009}. In this letter we report our experimental study of DD on an ensemble of cold atoms, where the fluctuations arise from velocity-changing elastic collisions in inhomogeneous fields. Our decoupling scheme is, to the best of our knowledge, the first experimental demonstration of the so called `Eulerian decoupling', i.e. a continuous robust version of DD \cite{PhysRevLett.90.037901}.

Cold atomic ensembles are promising candidates for quantum memory devices, which are the key component in modern quantum communication network architecture \cite{Duan2001}. Their main advantage is the high optical depth which was shown to be the main factor influencing the overall storage and retrieval efficiency \cite{gorshkov2007}. On the other hand, a large optical depth usually implies a fast collision rate, which leads to a rapid loss of coherence. Although many of the building blocks of quantum networks have been demonstrated experimentally using cold atomic ensembles \cite{Kuzmich2003,Chou2005,Yuan2008}, a short coherence time is a severe limitation. Recently, a record coherence time for quantum memory of several milliseconds \cite{Zhao2009,Zhao_R_2009} and storage time of $240ms$ for classical light storage in a Mott-insulator \cite{Schnorrberger2009} were reported. Nevertheless, the optical depth in these experiments was around $5$ which limits the overall efficiency to $\sim50\%$, compared to more than $95\%$ for $OD>100$.

An ensemble of two-level systems is usually subjected to two kinds of fluctuating fields that lead to dephasing: time and space dependent. Static inhomogeneities can be overcome completely by a single population inverting pulse ($\pi$ pulse) in the middle of the experiment - the celebrated coherence echo technique \cite{Hahn1950}. Time dependent fluctuations, on the other hand, pose a long-standing challenge, and their effect is usually irreversible. For atoms trapped in a red-detuned far off resonance laser beam there are several sources for inhomogeneous energy splitting between the internal states: differential ac \emph{Stark} shift due to a difference in the detuning for each internal state \cite{grimm2000}, density dependent interaction shift \cite{Harber2002} and inhomogeneity of externally applied fields. Coherence spin echo with cold atoms at low densities was demonstrated in \cite{Andersen2003,PhysRevLett.91.213002,PhysRevA.70.013405,Oblak2008}. The technique relies on the assumption that each atom is affected by the same averaged fields before and after the echo pulse. This assumption holds if the trajectory of each atom is bounded in some area in phase space, as is the case for Gaussian shaped traps but not, for example, for traps with underlying chaotic dynamics \cite{Andersen2004,Andersen2006_2}. Velocity-changing  elastic collisions randomize the atomic trajectories in the trap and hence are expected to prevent a coherence echo at high densities \cite{collisions_comment}. In addition to velocity-changing collisions, interatomic interactions contribute as a mean field energy shift, but if the density distribution is in equilibrium this has an effect of a static spatial inhomogeneity \cite{Harber2002}.

Our DD scheme reduces decoherence induced by collisions by applying a continuous pulse which induces population flipping between the two internal states (Rabi oscillations). Let us first consider a simpler case where we give a series of instantaneous $\pi$ pulses. If the time between the pulses is much shorter than the mean time between collisions, then in most cases there are no collisions during the echo period, and dephasing will be eliminated. Only pulses between which a collision occurs do not average out the inhomogeneous broadening, accumulate a phase and contribute to decoherence. We denote by $t_e$ the period between echo pulses, by $\Gamma_{col}$ the collision rate and by $\gamma_0$ the inhomogeneous broadening dephasing rate, {then} the decoherence factor for every $\Gamma_{col}^{-1}$ duration can be at most $t_e\gamma_0$, and the coherence decay rate can be written as $\gamma_{c}=\Gamma_{col}t_e\gamma_0$. Going to the continuous echo case, we substitute in the previous analysis $t_e=\pi\cdot \Omega^{-1}$, where we denote the Rabi frequency by $\Omega$, and find the following scaling law for the coherence time \cite{model_comment}
\begin{equation}\label{coherence_time_equation}
\tau_c \sim \gamma_0^{-1} \Gamma_{col}^{-1}\Omega/\pi \ \ .
\end{equation}
Notice that this derivation is only valid for $\Omega/\pi\gg \Gamma_{col}$.

\begin{figure}
    \begin{center}
    \includegraphics[width=7.5cm]{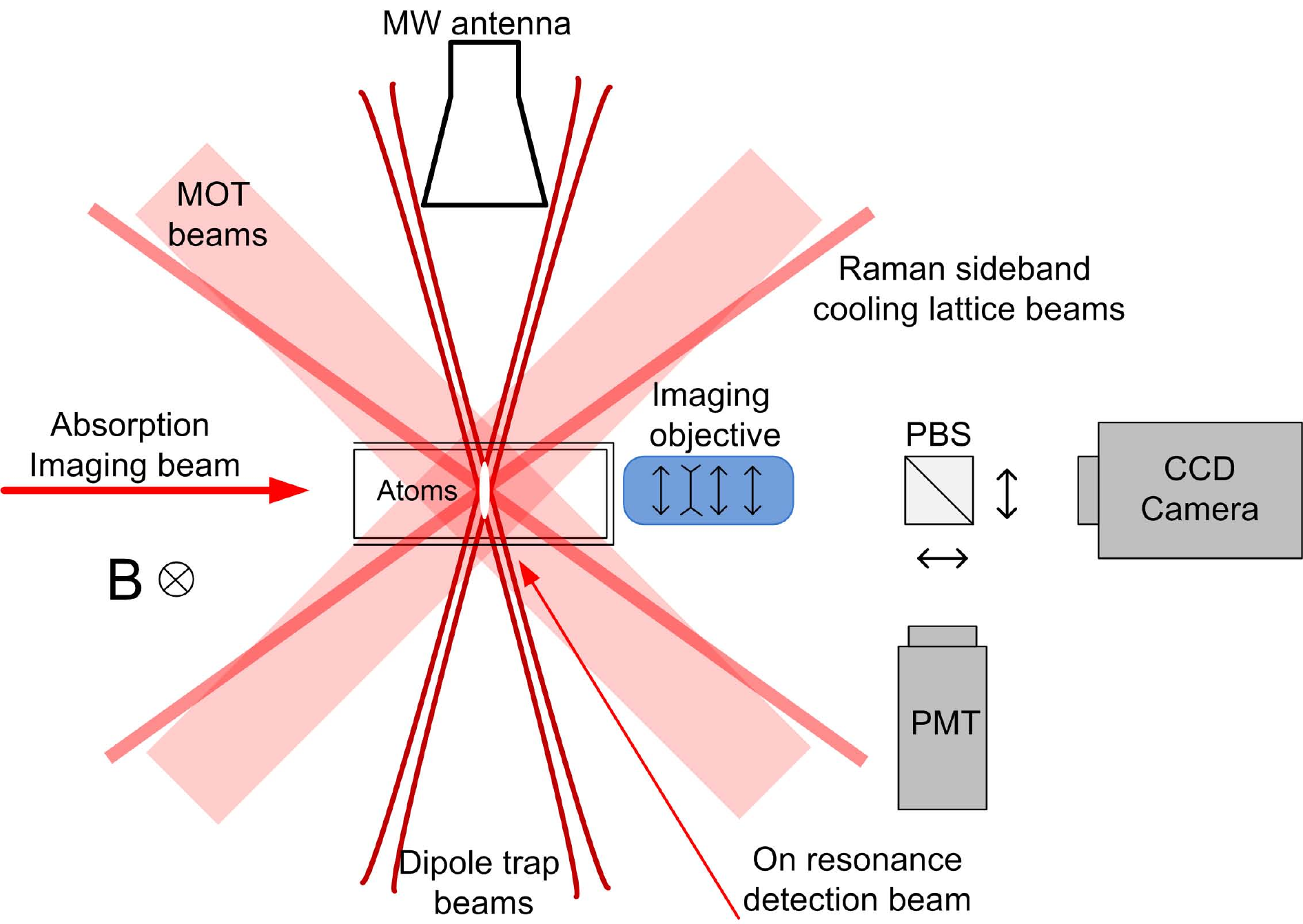}
    \end{center}\caption{The experimental setup. See description in text.}\label{experimental_setup}
\end{figure}

The experimental setup is depicted in \fig{\ref{experimental_setup}}. Initially $\sim10^9$ \Rb atoms are trapped and cooled in a magneto-optical trap (MOT), and further cooled by Sisyphus and Raman sideband cooling techniques \cite{Kerman2000}. At this point there are $\sim10^8$ atoms with a phase space density of $10^{-3}$, $80\%$ of them populating $\ket{5^2S_{1/2},F=1;m_f=1}$ and the rest populating $\ket{5^2S_{1/2},F=1;m_f=0}$. The atoms are loaded into a dipole trap created by two crossing beams at an angle of $28^\circ$, creating an oval trap with an aspect ratio of 1:3.9. The laser beams originate from a single frequency Ytterbium fiber laser at $1064$nm, and they have orthogonal polarizations and differ in frequency by $120$MHz to eliminate standing waves. The spontaneous scattering rate is less than $1s^{-1}$, and the trap lifetime is better than $5s$, both much longer than the relevant timescales in the experiment. The detection is done by either absorption imaging using a CCD camera or fluorescence detection using {a} photomultiplier tube. Using absorption imaging we can measure the density distribution of the atoms, and infer their temperature using time of flight technique. From measurements of the total number of atoms, temperature and trap oscillations frequencies we can determine the density and collision rate of the trapped cloud. Unless otherwise stated, the experiments are done with {a beams waist} of $50\mu m$ and total power of $\sim 6.4W$. We measure the radial oscillation frequency to be $2\pi \cdot 943$Hz. Evaporation is performed to remove atoms trapped in areas other than the crossing region. The final conditions are: total number of atoms $N=200,000$, temperature $T=22\mu K$, maximal density $\rho=7\cdot 10^{12} cm^{-3}$, and average collision rate $\Gamma_{col}=175 s^{-1}$  \cite{Wu1997}. The calculated optical depth for the long axis is $120$.

Interrogation and manipulation are done on the two internal levels: $\ket{1}=\ket{5^2S_{1/2},F=1;m_f=0}$ and $\ket{2}=\ket{5^2S_{1/2},F=2;m_f=0}$, which are to a first order Zeeman insensitive. We apply a magnetic field of $2.4G$ after the atoms are loaded into the dipole trap to lift the degeneracy of the $m\neq 0$ states. {A} microwave (MW) field is radiated to the atoms by a standard horn antenna, and due to magnetic dipole interaction with the atoms it induces coherent transitions between the two states. The antenna is driven by an amplified oscillator locked to atomic standard whose frequency and power we control by a computer. We measure the average atomic states population by means of {a} normalized state detection scheme \cite{Khaykovich2000}. We calibrate the transition frequency between the two internal states by inducing a long $100ms$ MW pulse whose power is adjusted to induce a population flip between the states ($\pi$ pulse). The transition frequency was measured to be $f_0=2\pi\cdot 6,834,685,665$Hz. All MW frequency detunings given hereafter will be relative to this value. Another important characterization of the setup is the spin relaxation time ($T_1$) which fundamentally limits the attainable coherence time. We measure the relaxation time by applying a $\pi$ pulse and after some time another $\pi$ pulse. $m$-changing spin exchange collisions in each hyperfine manifold cause depopulation of $\ket{2}$. We have measured a $1/e$ decay time of $T_1=317ms$ which is longer than all other relevant timescales in the experiment.

\begin{figure}
    \begin{center}
    \includegraphics[width=8.5cm,height=5cm]{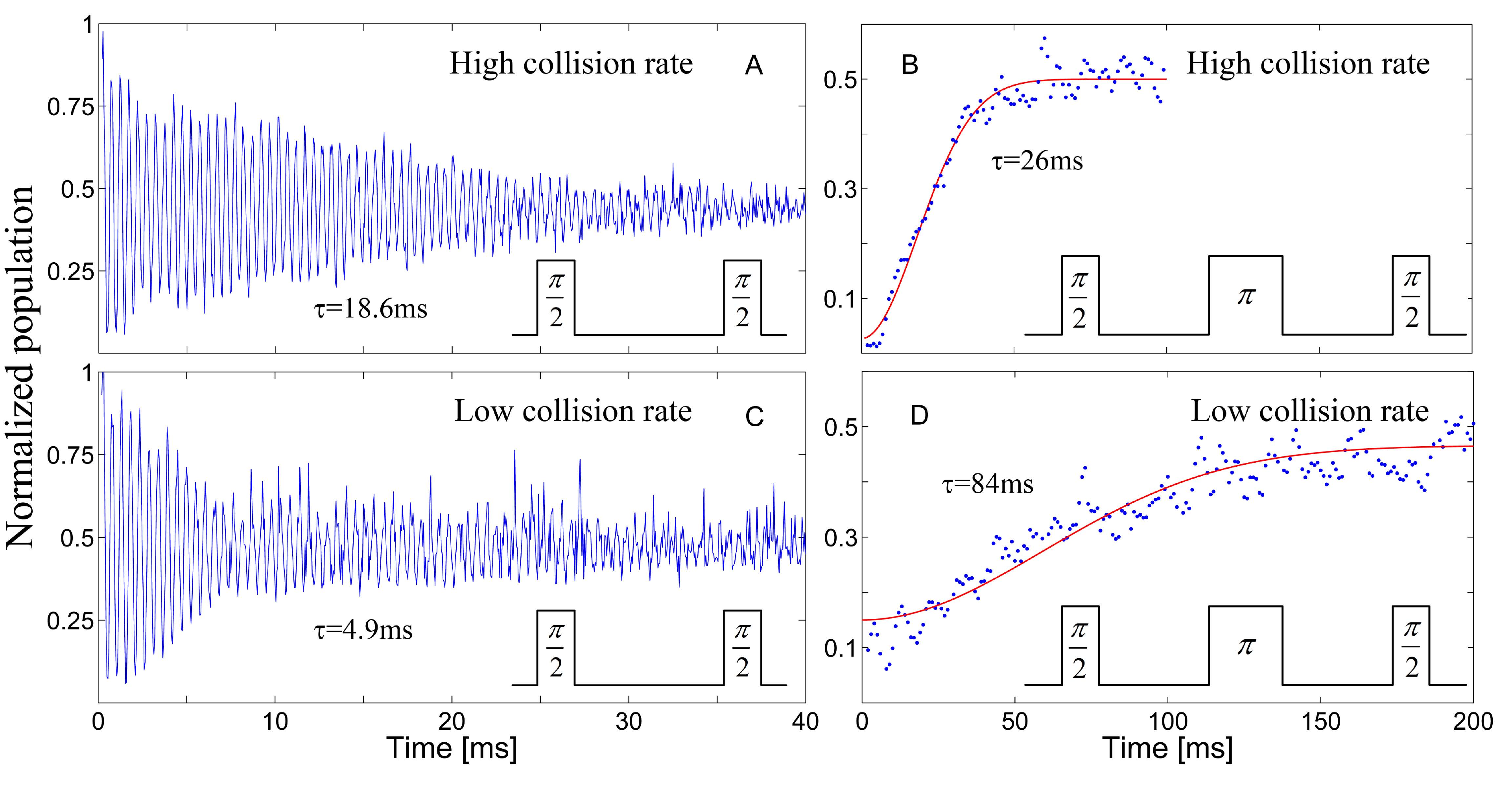}
    \end{center}\caption{Graphs A and C are Ramsey measurements and graphs B and D are echo measurements. Graphs A and B were done with four times the collision rate of graphs C and D. The conditions for experiments A and B were the ones already mentioned in the text, whereas experiments C and D were done in a different trap waist and power yielding the following conditions: $\omega_{osc,r}=2\pi\cdot 300$Hz, $T=17.5\mu K$, $\rho=2\cdot10^{12}cm^{-3}$, $N=1.2\cdot10^6$ and $\Gamma_{col}=43s^{-1}$. Red lines are fits to the function $ae^{-t^2/\tau^2}+b$. {The} x-axis in the Ramsey experiments is the time between the two pulses, and in the {echo experiments} it is the time between the first $\pi/2$ pulse and $\pi$ pulse. {The} y-axis is the population at $\ket{2}$ normalized to $0.2$. In the echo experiments since all population is initially at $\ket{1}$, a value of zero means complete coherence echo, and a value of 0.5 means complete decoherence. The Ramsey (echo) experiments were done with {a} detuning of $\delta=2\pi\cdot 2$kHz ($\delta=0$). {The} $\pi$ pulse duration in all experiments is $\sim 100\mu s$.}\label{ramsey_vs_echo}
\end{figure}

We characterize the dephasing caused by inhomogeneous broadening by means of a Ramsey experiment \cite{Ramsey56}. We prepare a coherent superposition \mbox{$\ppsi=1/\sqrt{2}(\ket{1}+\ket{2})$} by applying a $\pi/2$ pulse with a detuning $\delta=2\pi\cdot 2$kHz, wait some time and then give a second $\pi/2$ pulse. The Ramsey fringe visibility as a function of time is a measure to the coherence of the atoms. We fit the fringe envelope to the function $a+b(1+0.95\cdot(t/\tau)^2)^{-3/2}$ \cite{kuhr:023406}, where $\tau$ is the $1/e$ decay time. The experimental results, presented in \fig{\ref{ramsey_vs_echo}A}, show a decay time of $\tau=18.6ms$. Calculation of the inhomogeneous broadening weighted by the thermal distribution of the atoms in the traps yields an expected $\tau=9ms$, which is shorter than the measured value. We attribute the increase in the experimental dephasing time to slower atomic phase spreading owing to the collisions \cite{Harber2002}. In \fig{\ref{ramsey_vs_echo}B} the results of an echo experiment are shown together with a fit yielding $\tau=26ms$, showing that the echo indeed fails after a few collision events on average as expected. The role of collisions is farther emphasized by repeating these experiments with a lower collision rate of $43s^{-1}$. The results of a Ramsey experiment yielding a decay time of $\tau=5.1ms$ is shown in \fig{\ref{ramsey_vs_echo}C} and the corresponding echo experiment shown in \fig{\ref{ramsey_vs_echo}D} yields a decay time of $\tau=84ms$ which is more then $16$ times longer. A small revival is noticeable in the Ramsey signal at $t=10ms$, and it originates from the interference of the main Ramsey signal with the signal of atoms trapped in the wings of the crossed dipole trap. These atoms experience a different average detuning which corresponds to half the trap depth, which in our case is $2\pi\cdot65$Hz.

\begin{figure}
    \begin{center}
    \includegraphics[width=8.5cm]{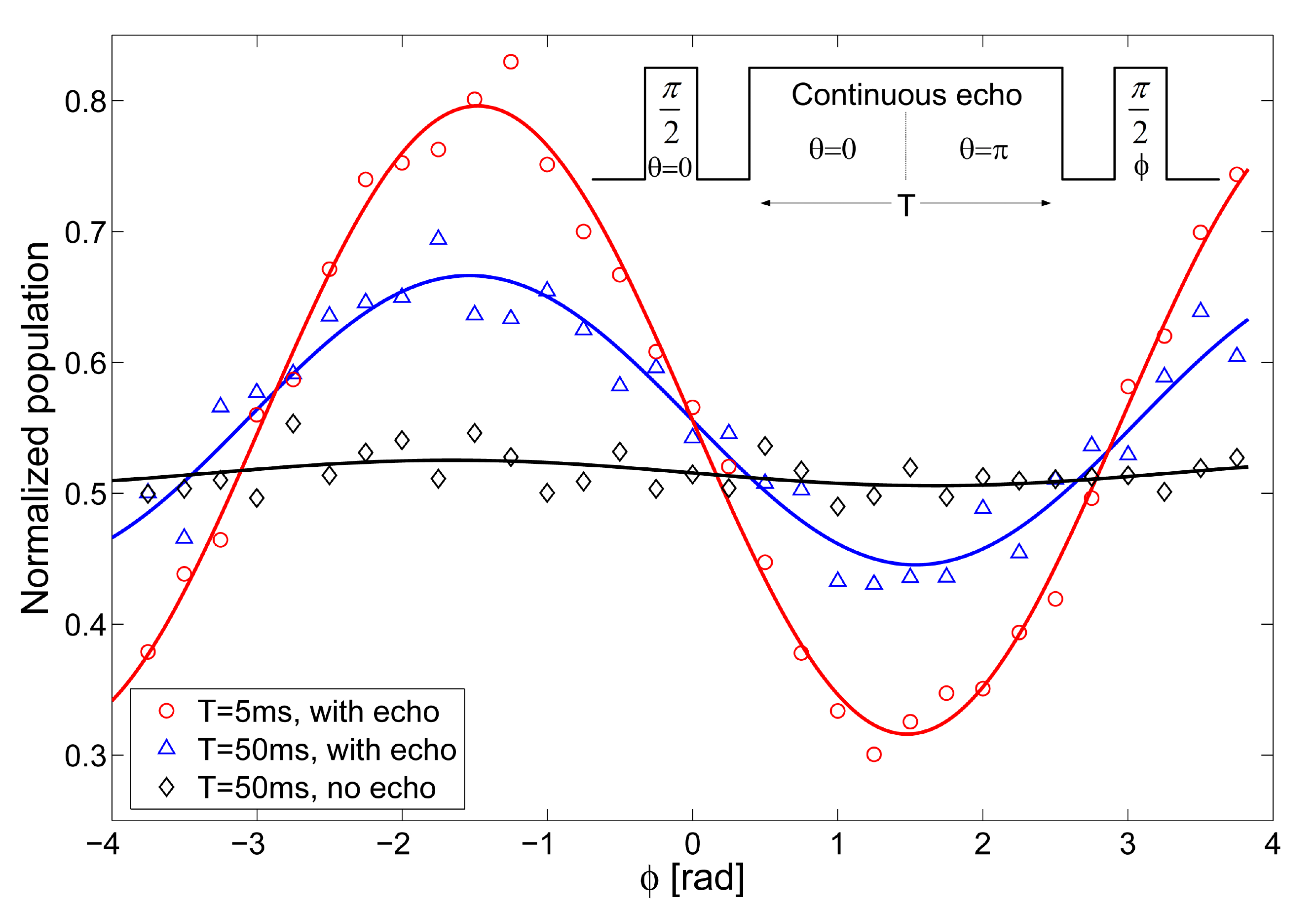}
    \end{center}\caption{Coherence contrast for $T=5ms$ (red), $T=50ms$ (blue) with a continuous echo pulse of $\Omega=2\pi\cdot 4333$Hz, and at $T=50ms$ without the pulse (black). {The} x-axis is the phase of the last pulse, and the y-axis is the normalized population at $\ket{2}$. We fit the data to the function $A\cos(kx+\pi/2)+B$, and find the amplitude A to be $0.24$, $0.11$ and $0.01$ for the red, blue and black datasets, respectively. The coherence time calculated from {these} values is $58ms$. The data is an average over 20 datasets for $T=5ms$ and 10 datasets for $T=50ms$, and due to this average there is a slight reduction in the contrast at $T=50ms$, which result in a slightly lower value for the coherence time compared to the results which are going to be presented later.}\label{ramsey_contrast_at_different_times}
\end{figure}

We now demonstrate the use of a continuous echo pulse to suppress decoherence due to collisions. We prepare an initial coherent superposition \mbox{$\ppsi=1/\sqrt{2}(\ket{1}+\ket{2})$} by applying an on-resonance $\pi/2$ pulse. We then switch on a continuous MW pulse with a Rabi frequency of $\Omega=2\pi\cdot 4333$Hz for a duration $T$, after which we give another $\pi/2$ pulse and scan its phase $\phi$. If the phase of the continuous pulse is kept constant then $T$ should be chosen such that $T=N\cdot 2\pi\Omega^{-1}$, where $N$ is an integer, and thus the final state is identical to the initial state. For large $T$ and $N$ this requirement results in a high sensitivity to several experimental parameters such as MW power and frequency and magnetic fields. Alternation of the phase of the MW radiation by $\pi$ every fixed duration $T_s$ can minimize this sensitivity. $T_s$ can be chosen freely as long as the total pulse duration $T$ is an even multiple of it. In the experiment presented here we switch the phase of the continuous echo pulse in the middle such that $T=2T_s$. The experimental results given in \fig{\ref{ramsey_contrast_at_different_times}} show the suppression of the decoherence of the state $\ppsi$. The results for $T=50ms$ show a significant coherent contrast compared to $T=5ms$, and as a reference we repeat the experiment without the continuous echo pulse (black graph) and observe complete decoherence. We have repeated these experiments with different initial states which can be prepared by changing the relative phase between the first and second MW pulses. We observe the suppression of decoherence for all initial states, but with different decay rates. The state $\ppsi$, in which the relative phase between the pulses in zero, gives the fastest decay rate, and therefore it constitutes the lower bound of the coherence time.

We now turn to quantitative characterization of the coherence time dependence on the Rabi frequency of the continuous echo pulse. In general, we want to prepare an initial coherent superposition \mbox{$\ppsi=1/\sqrt{2}(\ket{1}+\ket{2})$}, apply the continuous Rabi pulse for a changeable duration, and finally measure the ensemble average coherence. Measuring Rabi oscillations vs time is exactly such a measurement, since the first $\pi/2$ part of the oscillation prepares the state, afterwards the pulse continues and finally the fringe contrast after some time $t$ is a measure of the remaining coherence. An example of such a measurement for two Rabi frequencies is shown in \fig{\ref{comparison_of_raw_data_of_different_rabi_oscillations}}, where the slower decay rate of the envelope of the oscillations with higher Rabi frequency is evident. For $\Omega=2\pi\cdot 4800$Hz we observe a cumulative phase noise with time because of the reasons explained before, but for the present measurement it is not a problem since we are only interested in the decay of the oscillations envelope. In the analysis of the data we take the standard {deviation} of $24$ successive points and fit the result with an exponential decay function: $ae^{-t/\tau_c}+b$, where $\tau_c$ is the coherence time. It can be shown that such {an} analysis is not sensitive to the described phase noise. In the inset of \fig{\ref{coherence_time_vs_rabi_frequency}} we present the envelopes of three such measurements for three Rabi frequencies, together with their corresponding fits. In the main graph of \fig{\ref{coherence_time_vs_rabi_frequency}} we plot the measured coherence time as a function of the Rabi frequency. The expected linear dependency as derived in Eq.({\ref{coherence_time_equation}) can be clearly seen, from which we estimate the inhomogeneous dephasing to be $\gamma_0^{-1}=7.7ms$ in reasonable agreement with the calculated value of $9ms$. For the highest Rabi frequency we were able to apply we attained a coherence time of more than $100ms$.

\begin{figure}
    \begin{center}
    \includegraphics[width=8.5cm]{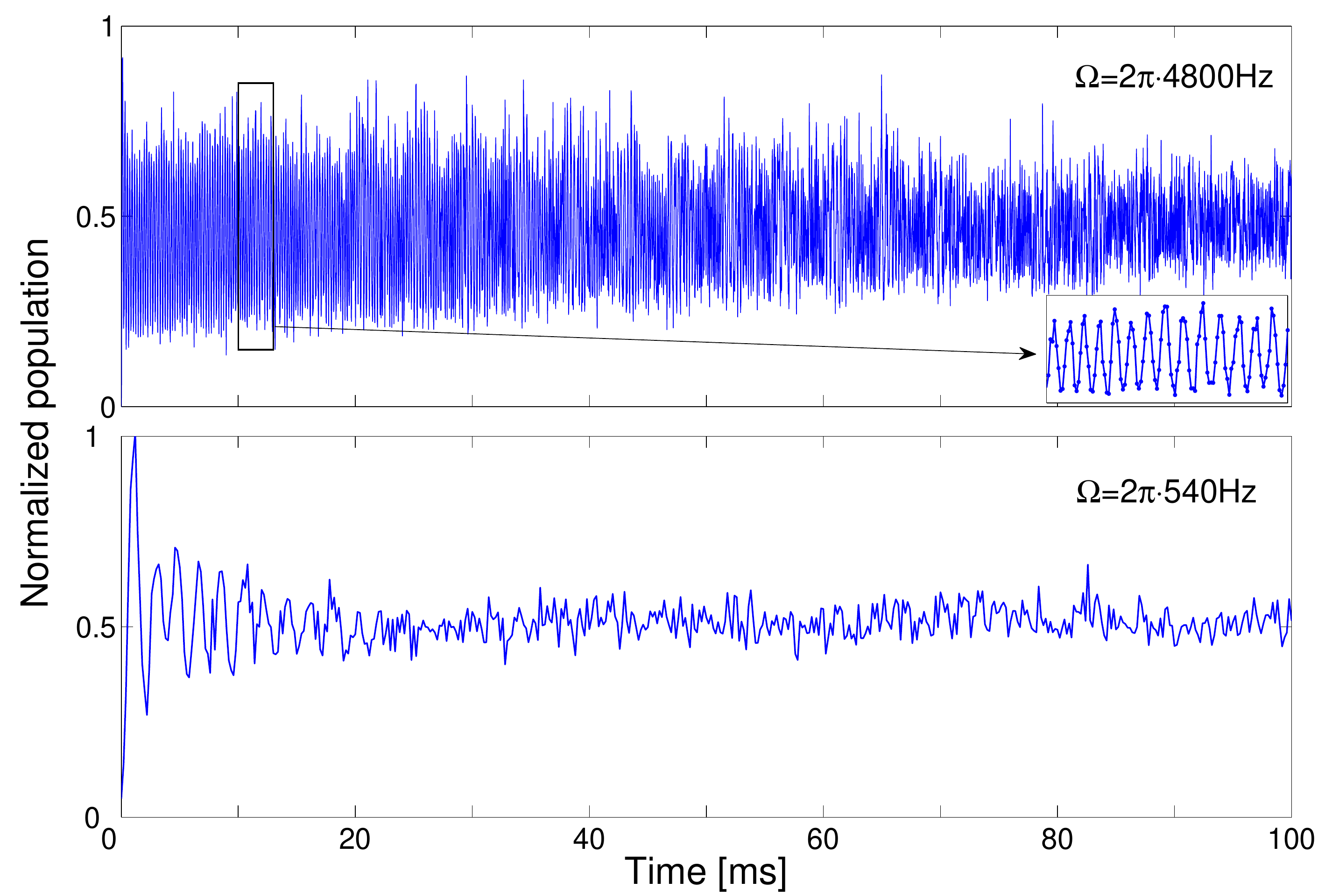}
    \end{center}\caption{Rabi oscillations with two different Rabi frequencies: $\Omega=2\pi\cdot 4800$Hz and $\Omega=2\pi\cdot 540$Hz for the lower and upper graphs, respectively. The x-axis is the duration of the MW pulse, and the y-axis is the normalized population at $\ket{2}$. The upper graph contains 4000 data points, with approximately 8 points per single Rabi oscillation. The inset is a zoom in on several such oscillations. datasets.}\label{comparison_of_raw_data_of_different_rabi_oscillations}
\end{figure}

\begin{figure}
    \begin{center}
    \includegraphics[width=8.5cm]{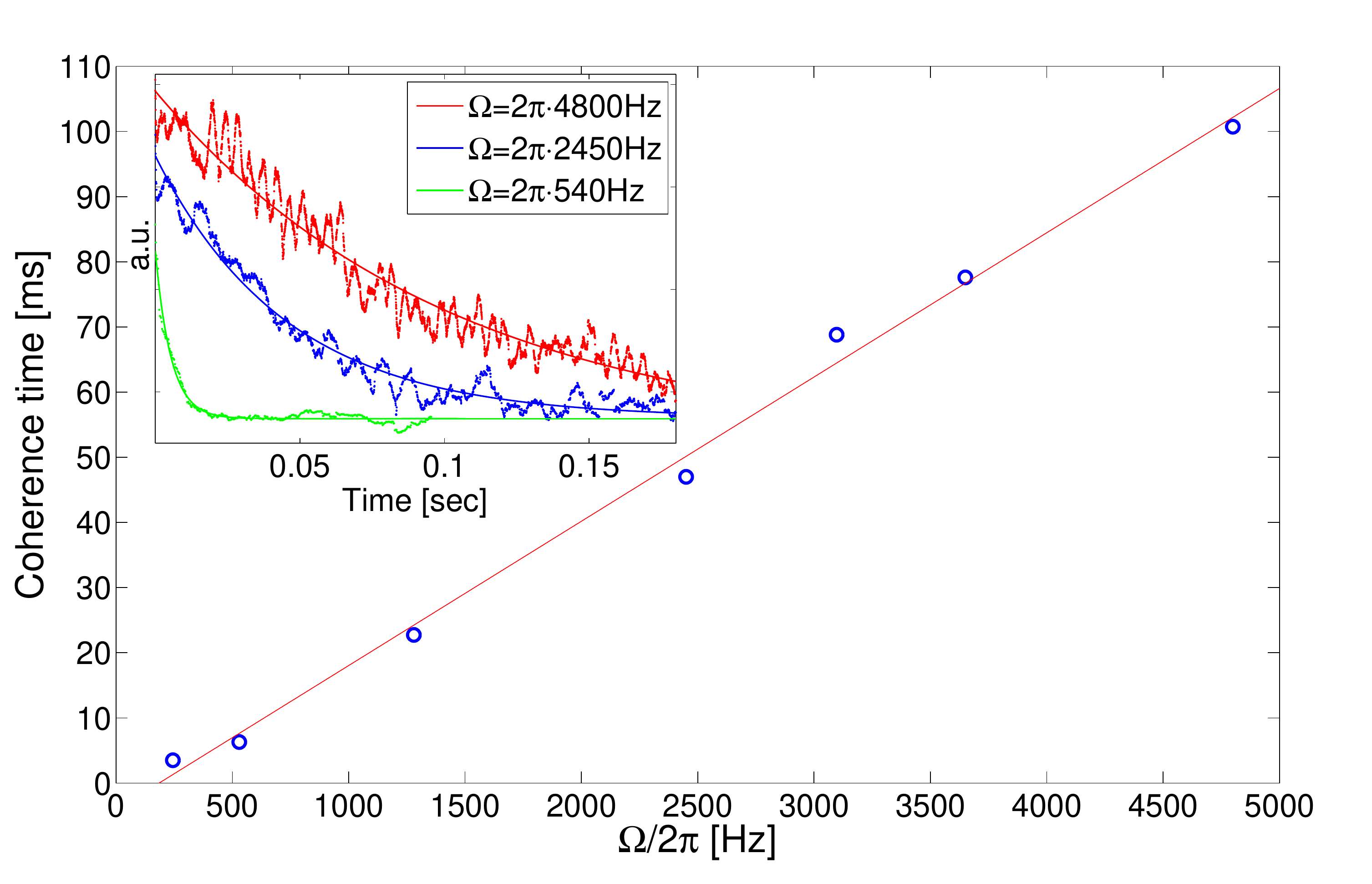}
    \end{center}\caption{Coherence time versus the Rabi frequency of the continuous echo pulse. The linear fit to the data yields: $\tau_c=0.022\cdot \Omega/2\pi-4.86$ (the linear fit only has a meaning for $\Omega\gg\Gamma_{col}$). The inset shows the envelopes of three Rabi frequencies together with their fits to the function $ae^{-t/\tau}+b$.}\label{coherence_time_vs_rabi_frequency}
\end{figure}

In summary, we have shown that elastic collisions in cold dense atomic ensembles randomize the atomic trajectories and cause the single echo technique to fail. Applying a continuous echo pulse increased the coherence time by a factor proportional to the ratio of the Rabi frequency to the average collision rate, giving a maximum coherence time of more than a $100ms$, where the limiting factor is the available MW power. Using higher microwave power and more sophisticated pulse sequences \cite{uhrig2007} it should be possible to increase the coherence time much more. Another promising prospect is performing light storage and quantum memory experiments incorporating our technique.

We thank Rami Pugatch for helpful discussions. We acknowledge the financial support of MINERVA, ISF, and DIP.


\begin{thebibliography}{30}
\bibitem[{col()}]{collisions_comment}
\bibinfo{note}{In this work we only deal with elastic collisions. Inelastic
  collisions cause mixing of internal spin states of the atoms, and lead to
  decoherence In our experiment the typical inelastic scattering rate is on the
  order of $1s^{-1}$, which is slower than the relevant timescales.}
\bibitem[{mod()}]{model_comment}
\bibinfo{note}{Note that since the effective Rabi frequency depend on the
  detuning there is a dephasing effect in Rabi oscillations due to
  inhomogeneous broadening, with a rate, however, that is much slower than the
  effects we demonstrate in this paper.}
\end{thebibliography}

\end{document}